 \documentclass[preprint2]{aastex}

\slugcomment{Accepted for publication in the Astrophysical Journal (Letters).}

\def\lsim{\mathrel{\rlap{\lower 4pt \hbox{\hskip 1pt $\sim$}}\raise 1pt \hbox
        {$<$}}}
\def\gsim{\mathrel{\rlap{\lower 4pt \hbox{\hskip 1pt $\sim$}}\raise 1pt \hbox
        {$>$}}}

\shorttitle{Precursors and Main-bursts of Gamma Ray Bursts
}

\shortauthors{Umeda, Tominaga}

\begin{document}
\title{Precursors and Main-bursts of Gamma Ray Bursts in a Hypernova Scenario
}

\author{\footnote{Department of Astronomy, University of Tokyo, Hongo, Bunkyo-ku, Tokyo
113-0033, Japan; umeda@astron.s.u-tokyo.ac.jp, tominaga@astron.s.u-tokyo.ac.jp, 
nomoto@astron.s.u-tokyo.ac.jp.
}
Hideyuki Umeda, $^1$Nozomu Tominaga, 
\footnote{Department of Earth Science and Astronomy, Graduate School of Arts and Science,
University of Tokyo, Komaba, Tokyo 153-8902, Japan; maeda@esa.c.u-tokyo.ac.jp}
Keiichi Maeda and $^1$Ken'ichi Nomoto}

%\affil{$^1$
%Research Center for the Early Universe, 
%University of Tokyo, Hongo, Bunkyo-ku,
%113-0033, Japan; umeda@astron.s.u-tokyo.ac.jp}

%\email{umeda@(;tominaga@;nomoto@)astron.s.u-tokyo.ac.jp; maeda@xxx.s.u-tokyo.ac.jp}

\begin{abstract}
 We investigate a "hypernova" model for gamma-ray bursts
(GRBs), i.e., massive C+O star model  
with relativistic jets. In this model, non-thermal precursors can be produced by
the "first" relativistic shell ejected from the star. Main GRBs are produced
behind the "first"-shell by the collisions of several relativistic shells.
They become visible to distant observers after the colliding region becomes optically thin. 
We examine six selected conditions using 
relativistic hydrodynamical simulations and simple analyses.
Interestingly, our simulations show that sub-relativistic $(v \sim 0.8c)$
jets from the central engine is sufficient to produce highly-relativistic
$(\Gamma > 100)$ shells. We find that the relativistic 
shells from such a star can reproduce observed GRBs with certain conditions.
Two conditions are especially important. One is the sufficiently long duration
of the central engine $ \gsim 100$ sec. The other is the 
existence of a dense-shell somewhere behind the "first"-shell. 
Under these conditions, both the existence and non-existence of
precursors, and long delay between precursors and main GRBs
can be explained.
\end{abstract}


\section{Introduction}

 Long Gamma-Ray Bursts (hereafter just GRBs) are likely produced by massive stars,
because they appear in star-forming regions (e.g., Vreeswijk et al. 2001; 
Gorosabel et al. 2003);
and some of them are followed by "bumps" in the afterglows that can be most convincingly
explained as bright supernovae (e.g., Reichart 1999; Bloom et al. 2002).
Most clear evidence was seen in GRB030329 whose afterglow turned into
a supernova (SN), SN2003dh (e.g., Stanek et al. 2003; Hjorth et al. 2003).
The spectrum of this SN showed that it was 
an energetic type Ic supernova, called "hypernova" (HN, e.g., Nomoto et al. 2004). 
Prototype of a HN is SN1998bw (Galama et al. 1998). The bumps in the
afterglows are also consistent with the light from HNe. 

 Since 1998, four HNe (SN1997ef, SN1998bw, SN2002ap, SN2003dh)
have been identified through the light curve and spectral modeling. 
The progenitors of these four are C+O stars of 5 to 14$M_\odot$
with explosion energies $ \sim 4 - 50 \times 10^{51}$ 
ergs (e.g., Nomoto et al. 2004).
Such C+O stars can be formed from massive stars with main-sequence masses
of $M \sim 20 -40M_\odot$ after losing outer envelopes by the strong
wind mass-loss or interaction with binary companion stars. 
We note that the existence of He layers may not affect on
the observed properties of the SNe.

 GRBs are produced from the shocks 
where highly relativistic shells collide with other shells 
forming internal shocks, or with circumstellar matter forming 
external shocks. Past studies have revealed that
the break in the afterglow light curve strongly suggests that 
the relativistic flow is 
in the form of jets with opening angle of $ 5 \sim 10^\circ$ 
(Frail et al. 2001). Variabilities
in the main-GRB favor the internal shock model than the external shock model,
though afterglows are well-explained by the external shocks.

 In this Letter we test a hypothesis that GRBs originate from 
massive C+O stars, and investigate whether the
time scales and observed properties of GRBs, including precursors, can be
consistently explained. 
We expect that adding He envelopes to the stars does not affect
the results much, because He stars have radius only few times 
larger than C+O stars. 
There are some previous works that considered the ejection of 
a relativistic jet from a massive star, but they mainly focused on
the jet formation in the magnetohydrodynamical processes 
(Mizuno et al. 2004; Proga et al. 2003) or the jet propagation
inside star and relatively soon after the breaking out of the
star (Aloy et al. 1999, 2002; Zhang, Woosley, \& Heger 2004),
and have not discussed much about the properties of GRBs themselves.

We discuss six conditions for successful models of GRBs 
by showing some results of numerical simulations and simple analyses 
to lead to general conclusions and to clarify remained issues. 
More detailed models and unconsidered effects 
such as rotation, MHD effects, and the effects 
of accompanied HNe are further explored in the following papers.

\section{Hydrodynamical Models}

 We have developed a multi-dimensional
special relativistic hydrodynamic code (Tominaga et al. 2005).
In this Letter we simulate relativistic jets from a 
14$M_\odot$ C+O star, embedded in the circumstellar matter (CSM)
with a density structure $r^{-2}$. 
The stellar radius is about $R_0 = 3.5 \times 10^{10}$ cm and CSM 
density at $r=10^{11}$ cm is
$10^{-11}$g cm$^{-3}$. The jet is initiated at $r= 2\times 10^9$ cm 
with opening angle of 15$^\circ$. At this location, the energy is 
injected mostly as thermal energy ($E_{\rm th}/E_{\rm kin} \simeq 40$)
with $v = 0.8 c$ and $\dot E \simeq 5\times 10^{51}$
erg s$^{-1}$ and the jet duration time, $\Delta t_{\rm jet} = 9$ sec.
In this model the total energy of $4.5\times 10^{52}$ erg is 
injected as a jet.  

 Here we show only the results of 2D calculations with axisymmetry
and with relatively coarse meshes, because the purpose of this Letter is to
make qualitative discussions. The adopted mesh size is
$3000 \times 45$ meshes with more meshes in the radial direction
in Eulerian spherical coordinate. This mesh size is sufficient to estimate 
the maximum Lorentz factors, $\Gamma$, within the error of $10\%$.
Higher resolution results as well as other parameter dependencies,
will be explored in the following papers (Tominaga et al. 2005). 

 The injected sub-relativistic jet is accelerated up tp $\Gamma \sim 50$
before breaking out of the star by the conversion from thermal to
kinetic energies. We note that the jet injection has to continue at 
least until the shock front breaks out of the star. Otherwise the
shock is strongly decelerated by the outer envelope of the star.
The evolution of the peak $\Gamma$ is shown in Figure 1a after $t=9$ sec
when the jet injection is terminated. This figure shows that the
peak $\Gamma$ continues to increase to $\Gamma \sim 160$ until
$t \sim 200$ sec because of the thermal to the kinetic energy conversion,
or thermal expansion.
In this model, $92\%$ of the total energy is contained within
9$^\circ$ from the jet-axis. 
The highest Mach number in the jet is 
about 10 that is achieved near the shock front of the jet.
Our calculations are similar to those in Zhang et al. (2004).
They assumed 
narrower and higher velocity jet injection, smaller 
$E_{\rm dot}$ and $E_{\rm tot}$ than our models,
but the results are qualitatively consistent. We can adopt a lower
velocity for the injected jet to obtain highly relativistic jet,
because our injected jet is thermal energy dominant.

\section{Six conditions to satisfy}

 Here, we will examine the following six conditions for a successful GRB 
model to satisfy.
(i) Can highly relativistic jets ($\Gamma > 100$) with sufficiently large
energy $E > 10^{51}$ erg be produced? 
(ii) Where should GRBs be produced? Is the GRB producing region optically thin?
(iii) Can the time scale of 
variability, $t_{\rm vari} \sim 1-20$sec be reproduced? 
(iv) Can the duration, $t_{\rm dur} \sim 100$sec of main-GRBs be reproduced? 
(v) Can the generation of precursors be reproduced? 
(vi) Can the long delay time between the
precursor and main-GRBs, $t_{\rm del} \sim 0 - 200$ sec be reproduced? 

(i) The GRB
spectra are non-thermal and considered to be produced by synchrotron 
radiation. According to the standard fireball model, 
relativistic shells with $\Gamma \gsim 100$ 
may produce such spectra (e.g., Piran 2005). 

 The relativistic shell must have sufficiently large energy to explain
the observed flux. For some gamma-ray bursts the estimated isotropic energy
is as large as $\sim 10^{54}$ erg. However, in the jet model the estimated 
energy can be as low as $\sim 10^{51}$erg. Therefore, the jet
should have energy at least $\sim 10^{51}$erg, though more than
$\sim 10^{52}$erg might be better considering energy convergent efficiency 
from the kinetic energy to radiation.

There are mainly two ways to eject relativistic materials from massive
stars. One is to eject point like masses with relatively high-density and 
small cross sections, known as the cannon-ball model (Dar \& Rujula 2004).
The other is narrow continuous relativistic jets that last at least 
until the front end of the jets breaks out of the star (Aloy et al. 1999; 
Zhang etal. 2004).
The former type objects cannot be significantly accelerated after
the ejection, thus highly relativistic matter with sufficiently large
density have to be produced near the central engine (Umeda 2000).
Since very little is known for the central engine, we can not exclude
such a possibility, but the latter type probably has less obstacles
because even sub-relativistic jets ($v \sim 0.8c$ near the center) 
are sufficient to produce highly relativistic jets (Fig.1).
As shown in Figure 1, jets from a C+O star can have sufficiently
large $\Gamma$ and $E$.

(ii)  We have shown that jets from a massive C+O star could potentially produce GRBs.
However the particular model in Figure 1 cannot reproduce the observed GRBs
because it is the single smooth jet and cannot produce variability
unless CSM is quite inhomogeneous. 
It is usually considered that the variability in GRBs is explained 
by the internal shocks produced
by the collisions of several shells. Such shells
might be produced by the variable activity of the central engine.
Then, several relativistic shells with various
$\Gamma$ may collide to produce GRBs through the internal shocks. 
Other mechanisms to produce several shocks have also been proposed
in the literature. For example, Aloy et al. (2002)  
suggested that a shear-driven instability leads to rapid fluctuations
that produce the internal shocks.
Although further study is certainly needed, we may say that the C+O stars with relativistic
jets are quite promising systems to explain the observed variability of GRBs.

 We assume  collisions of relativistic shells 
produce main GRBs. Since
the colliding regions are located behind the first relativistic shell,
the GRBs are not observable if the first shell is optically thick in gamma-rays. 
In Figure 1a we show the gamma-rays photosphere, where
the gamma-ray opacity is assumed to be 0.03 cm$^2$ g$^{-1}$.
This figure shows that around $t=1600$ sec, the
photosphere passes through the "first" relativistic shock.
Therefore, GRBs are visible only after 1600 sec in the local time. 
Is this too long or sufficiently short? The duration 
of the main-bursts is typically larger than 10 sec
in the observer's frame, that means the duration is 
$\sim 2 \Gamma^2 \times 10$ sec $ \gsim 2\times 10^5$ sec in the local time.
Therefore, the optical thickness is not a problem.

(iii) The observed time scale of variability is typically $t_{\rm vari} \simeq 1-20$sec
in the observer's time. In our model the distance 
between the internal shocks determines
the time scale. Let $l$ be the mean distance between the
internal shocks in the local frame of the 
star, then the mean interval of two GRB peaks 
from these shells is $l/c$ for observers
(Fig.2). Therefore, the observations require that $ l \simeq c\times t_{\rm vari} \simeq 
(3-60)\times 10^{10}$cm. Again, whether
this is possible or not depends on the properties
of unknown central engine. Successful model should explain this length scale.
This $l$ has the
same order to the radius of the progenitor star and this 
fact may be suggestive.

(iv) The duration time of the GRBs is typically $t_{\rm dur} \sim 100$ sec. 
This requires that the region where the 
internal shocks distribute should extend to the distance 
$L \sim c\times t_{\rm dur} \sim 3\times 10^{12}$ cm in the local frame (Fig.2). 
Roughly speaking this $L$ should be related to the jet duration 
time $\Delta t_{\rm jet}$
by $L \sim c\times \Delta t_{\rm jet}$, however as described below 
this $L$ can be 
larger than $c\times \Delta t_{\rm jet}$ by a factor of a few, 
considering the 
deceleration of relativistic shells by the collisions with other denser shells.
Thus we may write this condition as 

\begin{equation}
 \Delta t_{\rm jet} \gsim t_{\rm dur} \sim 100  \rm{~sec}.
\end{equation}

 Without knowing the mechanism of the central engine, it is difficult to discuss
whether such a jet duration time is reasonable or not. However, similar situation
has been investigated in the "collapsar model" 
(MacFadyen, Woosley \& Heger 2001). They suggested a possibility
to create narrow jets by the unspecified processes, such as the
Blandford-Znajek (1977) mechanism and other MHD precesses (e.g., 
Blandford \& Payne 1982). 

 It is reasonable to assume that $\Delta t_{\rm jet}$ is governed by the
time scale of mass accretion onto the central black hole. In the collapsar model,
the accretion time scale is short, $\sim 10$sec, for
the Type I case, where the outgoing shock fails to be launched from the collapsed
iron core. On the other hand, it is relatively long, $\sim 30-3000$ sec,
for the Type II case, where the inner layers of the star initially move outward 
but lack adequate momentum to eject all the matter. Therefore, the Type II
collapsar model at least can satisfy the above constraint (1). 

 In the jet-like explosion model of hypernovae (Maeda \& Nomoto 2003), 
most materials are ejected along the broad "jets", while
large amount of matter may fall-back for other directions.
For the equatorial direction, 
the fallback time scale may be as long as a few hundred seconds
as in the Type II collapsar model.

(v) The precursors are important to constrain the
models of GRBs. However their properties are still uncertain 
mainly because different authors define them differently.
We follow the definition by Lazzati (2005).  
Most precursors have non-thermal spectrum and they have equal
softness or softer spectrum than the main-bursts (Lazzati 2005).
They contain a small fraction
$(0.1-1 \%)$ of the total event counts. The delay time between the precursors and
main-GRBs, $t_{\rm dur}$, is typically the order of hundred seconds.
This Lazzati's precursors may be different from the X-ray precursors,
that are by definition not necessarily separated events from the
main GRBs. If the softness varies during the main GRBs, earlier or later part
of the bursts may be identified as X-ray precursors or postcursors
in some definitions (e.g., Nakamura 1999, for a model to explain them).

 The existence of the Lazzati's precursors is not well established, bu if
they are confirmed, 
long delays and the non-thermal spectra are difficult to reconcile
with conventional GRB models. For example, if both events are produced
by the same engine, it implies that after the engine forms a precursor, the
activity is silent for 100 seconds and then main-activity restarts. 
This seems to be the unlikely case.
The first jet may generate a flash of lights when it breaks out the star (e.g.,
Ramirez-Ruiz, MacFadyen \& Lazzati 2002); but the flash should have a 
thermal spectra, being inconsistent with the observed thermal spectra.

 Nevertheless our model, which is basically a standard jet-type model, 
can explain both the non-thermal spectra and the long delay
as follows. As shown in Figure 1, the first jet is highly relativistic and
is traveling into the CSM. Such a jet can produces an external shock
and non-thermal gamma-rays (e.g., Piran 2005). 
We propose that this is the observed precursor. 

 The relatively soft spectra may be reproduced if $\Gamma$ is not too large,
although determining all other synchrotron parameters are not so simple.
The small event count is consistent with this model, because the external
shock colliding with smooth CSM is inefficient in emitting gamma rays.
%We note that this "first jet" 
%may or may not directory become the observed afterglow. This is because,
%faster shell may catch up the first one from behind before it is observed as an afterglow.

(vi) Now the last question is how the long delay is explained. The delay 
can be explained if
the location of the external shock is separated from the closest internal shock
by the distance of $c \times t_{\rm del}\sim 3\times 10^{12}$cm in the local frame 
(Fig.2 left-bottom). 
This condition appears to be
difficult to satisfy because as mentioned in (iii) the GRB variability requires that the typical
separation between two internal shocks is $ l \simeq (3-60)\times 10^{10}$cm.

 This apparent problem can be solved if we assume the presence of
a sub-relativistic dense-shells somewhere behind the first shell. 
Such dense-shells may be formed from a shear-driven instabilities.
Although the existence of relatively slow shells have not been 
studied much previously, some variations in the velocities of the
shells are necessary for the internal shock model to work.
This is because if all the shells are highly relativistic, 
the kinetic to thermal energy conversion by the collisions of
these shells will be quite inefficient (e.g., Piran 2005).

 Let us consider a relativistic shell with a mass $M$ and the Lorentz 
factor $\Gamma$ colliding with a relatively slow dense-shell. 
The relativistic shell will be decelerated to $\sim \Gamma/2$
when the shell collides with a mass of $\sim M/(2\Gamma)$. 
Therefore, the shell is decelerated to sub-relativistic
when the mass of the dense-shell is larger than $\sim M$. 
If the dense-shell has a mass much larger than $M$, relativistic shells are
almost completely stopped, then large amount of 
thermal energy is generated and may turn into gamma-ray emission.
%For example, a shell with $\Gamma \sim 100$ and energy 
%$10^{51}$ erg is stopped when it collides with a matter
%with mass significantly larger than $M \sim 10^{-5} M_\odot$.  

 If such a dense-shell is non-relativistic, then the distance 
between the shell and
the "first"-relativistic shell increases with time (see Figure 2 left). 
The main GRBs that occur behind the dense-shell will be visible only after 
the dense-shell becomes
optically thin at $t_{\rm opt}$. The delay time is then estimated by 
$t_{\rm del} \simeq l_{\rm opt}/c$, where $l_{\rm opt}$ is the distance between the
dense-shell and the "first"-shell at $t_{\rm opt}$.

 For the model in Figure 1, at $t=100$ sec, the "first"-shell is located 
at around radius $=94$, in the units of $c \times 1$ sec =1. Let the
position, or the distance from the center of the explosion,
of the shell at time $t$ be $x_1$, then it is approximately
$x_1 \simeq 94+ c\times(t-100)= t-6$.
Let us consider a dense-shell located at radius $= x_{20} (\sim x_1-l)$ 
at $t=100$ with radial velocity $\alpha (<1)$. The position of the
dense-shell at $t$ is written as $x_2= x_{20} +\alpha (t-100)$. In terms of these
quantities $t_{\rm del}$ is 
written as $ t_{\rm del} \simeq (x_1-x_2)/c = (1-\alpha)t_{\rm opt} -6+100\alpha-x_{20}$.
Here we assume that the velocity of the dense-shell is constant, but
it can increase with time by the interaction with surrounding material
and the collision with the relativistic matter from behind. For simplicity
let us assume that the shell has a constant velocity $\alpha$ until it
becomes relativistic at time $t_{\rm rel}$.
Then the above condition can be rewritten as

\begin{equation}
 t_{\rm del} \simeq (1-\alpha)t_2 -6+100\alpha-x_{20}, 
\end{equation}
where $t_2 = $min$(t_{\rm opt}, t_{\rm rel})$. For example, when $x_{20}=80$, 
$\alpha=0.85$, and $ t_{del}\gsim 100$, $t_2$ is constrained as
$t_2 \gsim 670$ sec.

 There is another constraint that has to be satisfied: if the jet duration
time, $\Delta t_{\rm jet}$, is too short, the collisions of shocks finish
before the GRBs become visible. Let $x_3$ be the position of the "last"
relativistic shell and suppose that $x_3 = x_{30}$ at $t=\Delta t_{\rm jet}$,
then $x_3(t) \simeq x_{30} + (t-\Delta t_{\rm jet})$. 
Here typically $x_{30}$ is $ x_{30} < c\times$10 sec.
This constraint is then written as $ x_3(t_2) < x_2(t_2)$, or,
$ \Delta t_{\rm jet} > x_{30} - x_{20} + (1-\alpha)t_2 + 100\alpha$.
Using equation (2), this becomes
$\Delta t_{\rm jet} \gsim x_{30}+t_{\rm del}+6.$
If $t_{\rm del}+ x_{30} \gsim 100$, this constraint is roughly the same as 
Equation (1).

 We note that some GRBs do not have any indications of precursors.
In our model this is simply understood as the very small $t_{\rm del}$.
$t_{\rm del}$ may be nearly zero if there is no slow dense-shell or
if the dense-shells are accelerated to relativistic speed quickly
by the collisions with following shells.
 
\section{Conclusion}

 We find that the relativistic 
jets from a C+O star can reproduce observed GRBs with certain conditions.
Two conditions are especially important. One is the sufficiently long duration
time for the jet, $\Delta t_{\rm jet} \gsim 100$ sec. In this Letter
we treat the central engine as a black box, thus being unable to answer if such a
condition is actually satisfied. However, our model may explain such duration
because HNe is likely to be an
aspherical explosion where large fall-back
may occur from equatorial direction possibly for a long time. 
We will investigate such explosions in future works.

 The other important condition is the
existence of a sufficiently massive sub-relativistic "dense-shell". 
The initial mass
and velocity of such a shell together with the sufficiently long
$\Delta t_{\rm jet}$ can produce long $t_{\rm del}$, as well as
short $t_{\rm del}$ from a relatively short time scale $t_{\rm vari} \simeq
1 - 20$ sec. This conclusion is robust because the existence of 
relativistic shells in front of the slow-dense-shell merge into the
first-shell without forming a strong internal shocks.
In our model, if two or more slow dense-shells are formed, 
main GRBs are split into two or more parts. This phenomena appear to
be seen in some of actual GRB events.
In order to answer how the dense-shells and other relativistic
shells with mean separation $l \sim 10^{11}$ cm
are formed, we at least need to perform higher resolution calculations
to study hydrodynamical instabilities. We also need to investigate the
effects of He-envelopes for the formation and distribution of those shells.
Other unconsidered effects such as magnetohydrodynamical effects 
may be critical for the instabilities. We leave such studies for 
future works.

%\end{document}

\clearpage

\begin{figure*}
\vspace{-4cm}
\epsscale{3.}
\plotone{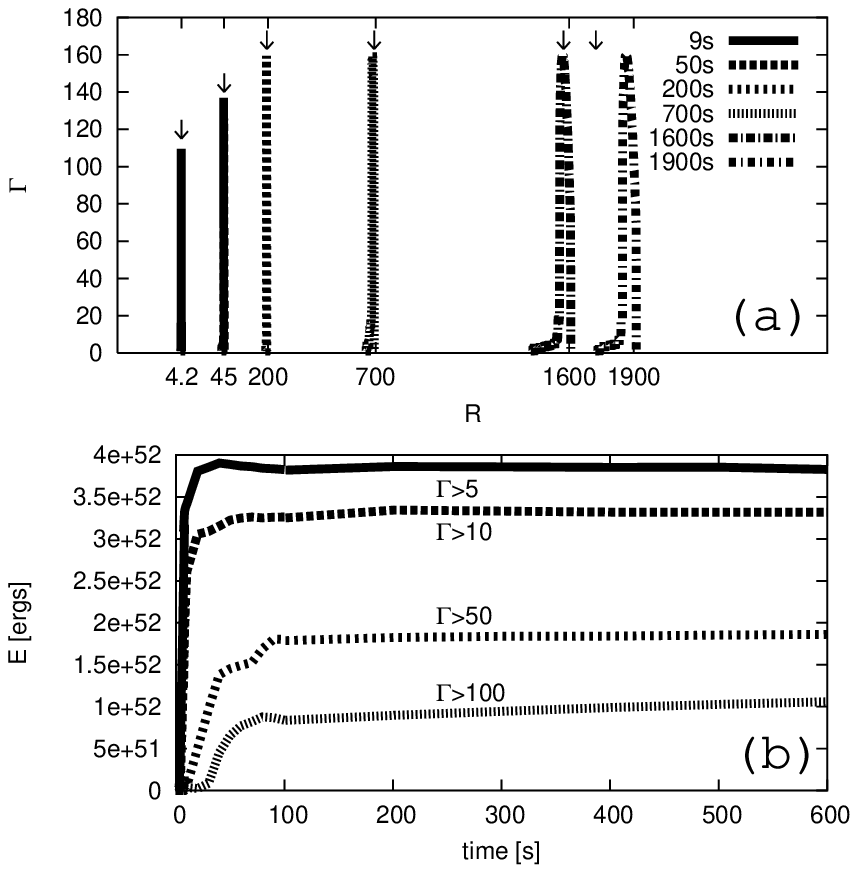}
%\plotone{RWLF.eps}
\caption{Distribution of $\Gamma$ along the jet axis for the
model in Section 2, plotted against 
the radius, $R$, in the units of $c\times 1$ sec =1 (Fig.1a).
The location of gamma-ray photosphere is shown by arrows.
The relativistic shock becomes optically thin for $t > \sim 1600$ sec.
The energy contained in matter
with $\Gamma$ greater than the indicated values is shown in
Figure 1b.
}
\end{figure*}

\clearpage

\begin{figure*}
\epsscale{2.}
%\plottwo{25z0e1A.ps}{25z0e20A.ps}
\plotone{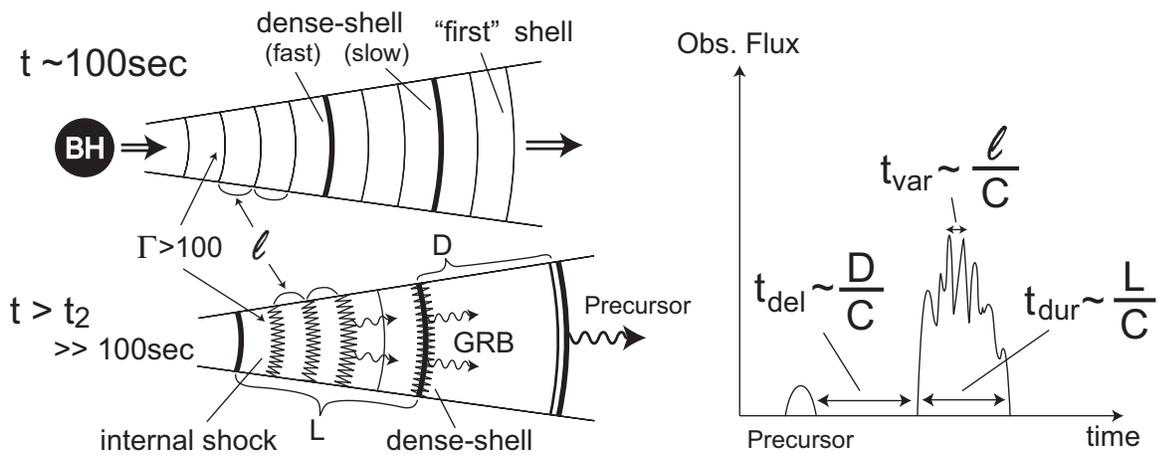}

\caption{Schematic pictures of our GRB model in the local frame (left),
and gamma-ray flux in the observer's frame (right). The distance between
the initially slow dense-shell and the front most shell, $'D'$, determines
the delay between the precursor and main-bursts.}
\end{figure*}

\end{document}